\documentclass[twocolumn,showpacs,aps,prl,letter,amsmath,amssymb,superscriptaddress]{revtex4-1}
\usepackage{bm,color,bbm}
\usepackage{mathtools,graphicx,natbib}

\usepackage[english]{babel}
\usepackage[utf8x]{inputenc}
\usepackage{float}

\usepackage{siunitx} % Required for alignment

\newcommand{\beq}{\begin{equation}}
\newcommand{\eeq}{\end{equation}}
\newcommand{\bqa}{\begin{eqnarray}}
\newcommand{\eqa}{\end{eqnarray}}

\newcommand{\red}{\color{black}}

\newcommand{\blk}{\color{black}}
\newcommand{\blu}{\color{black}}
\definecolor{maroon}{rgb}{0.7,0,0}

\definecolor{ngreen}{rgb}{0.3,0.7,0.3}

\definecolor{golden}{rgb}{0.8,0.6,0.1}

\usepackage{amsmath}
\usepackage{graphicx}
\usepackage{epstopdf}
\usepackage{amssymb}

\begin{document}

\title{\red Experimental Realization of \blk a Quantum Autoencoder: \\ The Compression of Qutrits via Machine Learning}

\author{Alex Pepper}
\affiliation{Centre for Quantum Dynamics, Griffith University, Brisbane, QLD 4111, Australia.}

\author{Nora Tischler}
\email{n.tischler@griffith.edu.au}
\affiliation{Centre for Quantum Dynamics, Griffith University, Brisbane, QLD 4111, Australia.}

\author{Geoff J. Pryde}
\email{g.pryde@griffith.edu.au}
\affiliation{Centre for Quantum Dynamics, Griffith University, Brisbane, QLD 4111, Australia.}

\begin{abstract}
With quantum resources a precious commodity, their efficient use is highly desirable. Quantum autoencoders have been proposed as a way to reduce quantum memory requirements. Generally, an autoencoder is a device that uses machine learning to compress inputs, that is, to represent the input data in a lower-dimensional space. Here, we experimentally realize a quantum autoencoder, which learns how to compress quantum data using a classical optimization routine. We demonstrate that when the inherent structure of the dataset allows lossless compression, our autoencoder reduces qutrits to qubits with low error levels. We also show that the device is able to perform with minimal prior information about the quantum data or physical system and is robust to perturbations during its optimization routine.

\end{abstract}

\maketitle

\paragraph{Introduction.---} 

Quantum technologies promise to provide us with advantages over their classical counterparts in a variety of tasks, including faster computation \cite{Grover1996}, secure communication \cite{Nielsen2010}, and increased measurement precision \cite{Giovannetti2011,Slussarenko2017}. However, they depend on quantum resources, for example quantum coherence, which can be challenging to produce, control, and preserve effectively. As such, quantum resources are precious, and devices that allow us to minimize the use of these resources are valuable. One such device is the quantum autoencoder \cite{Romero2017,Wan2017,Lamata2017,Steinbrecher2018}. 

\begin{figure}[b!]
	\centering
	\includegraphics[width=8.6cm]{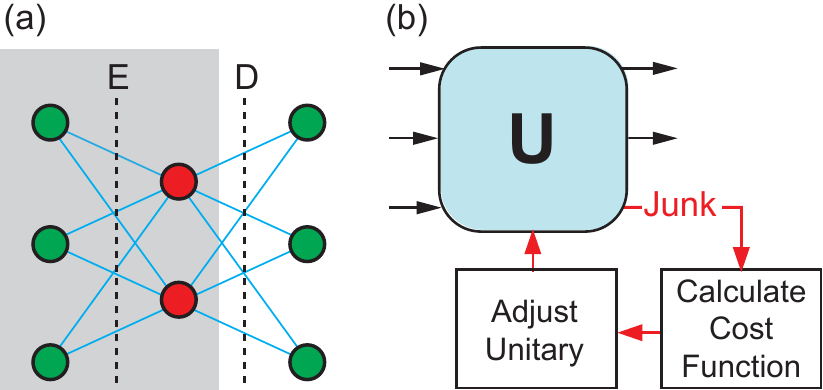}
	\caption{(a) The concept of an autoencoder. Through an encoding process (E), autoencoders represent data in a lower-dimensional space; if the compression is lossless, the original inputs can be perfectly recovered through a decoding process (D). 
(b) The scheme of our qudit-based autoencoder, equivalent to the gray shaded section in (a), for the case of a compression of qutrits to qubits. A unitary transformation, $U$, characterized by a set of parameters converts between three input modes and three output modes. The iterative training of the parameters aims to minimize the occupation probability of the third output mode, the ``junk'' mode, across a set of training input states. Lossless compression is achieved when the junk mode is unoccupied. Given that the encoding step is performed with a unitary transformation, the decoding step would simply be the inverse of the unitary, using a vacuum state in the third input mode.}
\label{fig:simple}
\end{figure}

An autoencoder uses machine learning to represent data in a lower-dimensional space, as illustrated in Fig.\ \ref{fig:simple}(a). Autoencoders for classical data form one of the core architectures in machine learning, and offer a range of tools for image processing and other applications \cite{Guo2016,Liu2017,Menshawy2018}. Models of autoencoders that compress \emph{quantum} data were recently proposed and theoretically studied in Refs.\ \cite{Romero2017,Wan2017,Lamata2017,Steinbrecher2018}\footnote{During the writing of this article, we became aware of another, parallel experimental implementation of a quantum autoencoder, reported in Ref.\  \cite{Ding2018}.}.\nocite{Ding2018}

Quantum data compression can benefit applications such as quantum simulation \cite{Romero2017,Thompson2018} and the communication and distributed computation between nodes in a quantum network \cite{Lamata2017,Steinbrecher2018}, by reducing requirements on quantum memory \cite{Romero2017,Rozema2014}, quantum communication channels \cite{Steinbrecher2018}, and the size of quantum gates \cite{Lamata2017,Ding2018}. Reversible, and therefore lossless, compression is possible if a set of quantum states does not span the full Hilbert space in which they are initially encoded. Some methods of compressing quantum data have been proposed and demonstrated previously \cite{Bartuskova2006b,Rozema2014}. In those methods, the compression was based on specific assumptions about properties of the quantum states, for example that a set of qubits be separable. The main advantage of autoencoders is that they do not rely on assumptions of this kind; instead of exploiting a fixed structure of the data, they are capable of learning the structure based on a training dataset. This ability to learn makes them versatile.

\begin{figure*}[t]
	\centering
	\includegraphics[width=17.2cm]{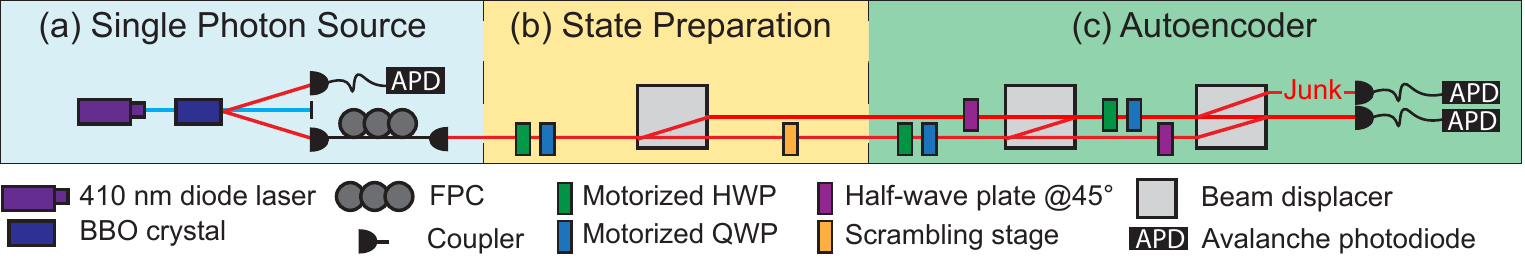}
	\caption{Experimental layout. (a) Single photon source. We use a 410 nm continuous-wave diode laser to pump a beta barium borate (BBO) crystal to produce photon pairs via type-I spontaneous parametric down-conversion. One photon is collected and detected to herald the second photon; this second photon is fiber coupled and passes through a fiber polarization controller (FPC). (b) State preparation. First, a qubit is encoded in the polarization state of the photon via a motorized half-wave plate (HWP) and quarter-wave plate (QWP). Then the photon passes through a polarizing beam displacer. In the lower spatial mode, a random but fixed birefringent element (a randomly oriented wave plate for a different wavelength) scrambles the polarization, mapping vertical polarization to some superposition of horizontal and vertical. The top spatial mode has matching optical elements (not shown) set to the optic axis, in order to match the optical path length to within the photon's coherence length. By changing the motorized wave plates, sets of qutrits are created in polarization and path modes in such a way that they are compressible into qubits. (c) Autoencoder. A $3\times 3$ unitary transformation with four free parameters is implemented via a series of wave plates, which perform $2\times 2$ unitaries, and beam displacers, which realize mode permutations. The training of the unitary transformation is performed through the rotation of the motorized half- and quarter-wave plates, using a gradient descent routine. The cost function is evaluated based on the measured photon occupation probability of the upper output mode, which contains only one polarization. }
\label{fig:setup}
\end{figure*}

Here, we propose and experimentally realize a simple quantum autoencoder scheme. Our device works on a similar principle to the theoretical model of Ref.\  \cite{Romero2017}, but with some notable differences. Whereas the proposal of Ref.\ \cite{Romero2017} concerns the compression of some number of qubits to fewer qubits, we pursue a dimensionality reduction in a qudit representation (see Fig.\ \ref{fig:simple}(b)). Consequently, the desired outputs of our device, as well as the definition and evaluation of the cost function, differ from the original model, as described later on. Generally, our scheme supports a compression of qudits to qunits, where $d>n$ \cite{SI}. We demonstrate it experimentally for the case of $d=3$, $n=2$, with a photonic device that reduces qutrits to qubits. One potential use of this type of compression lies in applications of quantum communication, for instance between nodes in a quantum network. In quantum communication, photons are a natural choice as quantum information carriers due to their high speed and strong coherence properties. Moreover, the output qubit from our device is encoded in the polarization degree of freedom, which is easily manipulated and can be transmitted over long distances \cite{Krenn2016}.

The combination of machine learning and quantum information processing is a growing research area, which aims to either draw on classical machine learning techniques to aid quantum information tasks, or utilize quantum information processing to speed up classical machine learning calculations \cite{Dunjko2018,Biamonte2017}. The quantum autoencoder belongs to the former category. Complementary to recent demonstrations of several classification tasks \cite{Li2015,Cai2015,Gao2018}, Hamiltonian learning \cite{Wang2017}, and the reconstruction of quantum states \cite{Yu2018}, the present work experimentally establishes the compression of quantum data as another use of quantum machine learning. 

In this work, we apply the quantum autoencoder to the task of lossless compression, for which we use families of qutrit states that are compressible to qubits. The device can be trained based on a few examples of the family of qutrit states, and can subsequently be tested with other qutrits from the family. Our autoencoder uses a classical machine learning algorithm, gradient descent, to optimize a unitary transformation for compressing the quantum states \cite{SI}. The automated experimental unitary optimization circumvents the need to obtain a classical description of the training  states, externally design the appropriate unitary, and carry out a full characterization of the optical elements. As an additional benefit, the device is also robust to disturbances during its optimization routine, as discussed later. These advantages of optimizing directly based on experimental data are akin to previous observations in other systems \cite{Frank2017,Gao2018}.

\paragraph{Experimental scheme.---}

Our autoencoder consists of a $3\times 3$ unitary transformation with four free parameters, along with a feedback mechanism that is based on measurements of one of the output modes. The setup of our photonic implementation is depicted in Fig.\ \ref{fig:setup}. \red The input qutrits are encoded as superpositions over three optical modes: one spatial mode supporting two polarization modes, and another spatial mode with a fixed polarization. \blk The transformation is implemented as a sequence of $2\times 2$ unitaries, each realized by a set of half- and quarter-wave plates, while mode permutations are realized by beam displacers and half-wave plates set to $45^{\circ}$ \cite{Laing2011,Martin-Lopez2014} \footnote{An implementation of an arbitrary unitary transformation as a sequence of $2\times 2$ unitaries is always possible, as established in Refs.\ \cite{Reck1994,Laing2011,Martin-Lopez2014}. A general unitary would additionally require a half- and quarter-wave plate in the lower spatial mode of Fig.\ \ref{fig:setup}, as well as single-mode phase elements, before the detectors. However, these elements are unnecessary for our purposes, since they do not affect our cost function.}. \nocite{Reck1994} Such an implementation using joint polarization and spatial modes provides high stability, \blu because all the spatial modes enter and exit the beam displacers through the same facets, and it allows \blk a simple way of controlling the unitary transformation.

Our quantum states are encoded in single photons. \blu In principle, the training process could also be performed with coherent states instead of single photons. However, in applications of the quantum autoencoder within quantum technologies, the training states are more likely to be available in the form of single photons. \blk Given a set of training states, the device performs a unitary transformation with the aim to map all of the training qutrits onto qubits. The desired mapping is achieved when the third output mode, which we will refer to as the ``junk'' mode,  is unused; that is, when the photon cannot be found in that mode (see Fig.\ \ref{fig:simple}(b)). In that case, the mode can be discarded without any effect, leaving the outputs in a qubit space. By contrast, whenever the compression to the qubit output is imperfect, there is a nonzero probability of finding photons in the junk output, and this can be interpreted as a measure of error. 
The occupation probability of the junk mode, $P_{\mathrm{j}}$, quantifies the compression performance twofold: (i) the success probability of the encoding process is $1-P_{\mathrm{j}}$, and (ii) the fidelity between the given input state and the output after an encoding-decoding sequence is likewise $1-P_{\mathrm{j}}$ \cite{SI}.
Therefore,  the goal of the training process is to minimize the occupation probability of the junk mode across the training states, ideally to zero. In our experiment, we define the cost function as the \blu average \blk junk mode occupation probability over the different training states. We probe the junk output, calculate the cost function, and use a classical gradient descent optimization routine to adjust the unitary \blu \cite{SI}\blk.

In order to prepare our input states, polarization qubits are created with a single photon source paired with a half- and quarter-wave plate. We then map these states onto qutrits by using a beam displacer and scrambling the polarization of one spatial output mode. Although the resulting qutrits occupy all three modes, by construction, they lie within a subspace that allows them to be compressed to qubits. We obtain the training states, as well as further states to test the performance of the autoencoder, by varying the half- and quarter-wave plate settings in the state preparation. 

\begin{figure}[tb]
	\centering
	\includegraphics[width=8.6cm]{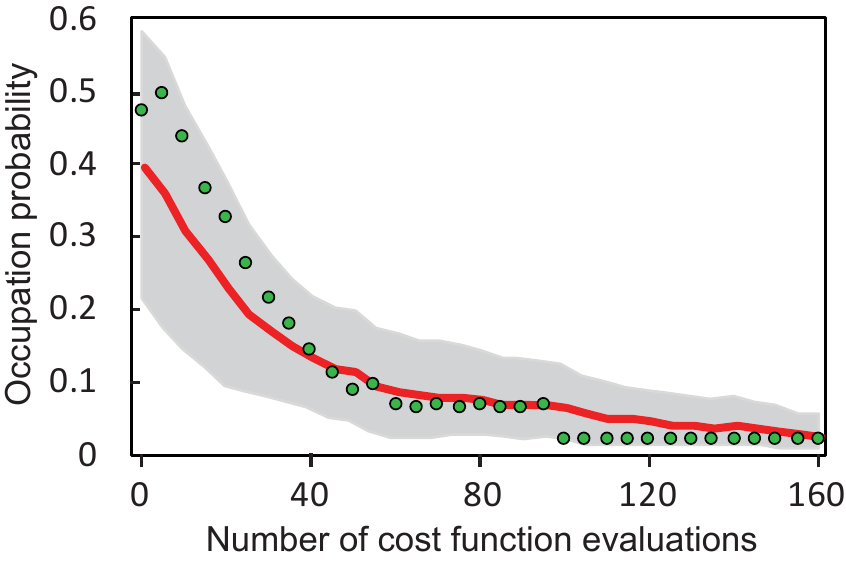}
	\caption{Training starting from different initializations of the unitary. Here, we used two fixed, randomly selected training states to perform the optimization routine. The \blu cost function, which is the \blk occupation probability of the junk mode averaged over the two training states, is plotted against the number of cost function evaluations. The optimization was carried out twenty times, each time starting with a different randomly initialized unitary. To save training time, a training run was terminated once the average occupation probability reached a threshold of 0.02, and the average occupation probability for that training run was thereafter set to the last measured value. The red line shows the mean values of the 20 training runs, with the gray shaded area indicating $\pm $ one standard deviation of the results.  We observe an average occupation probability of 0.03±0.03 after 160 cost function evaluations. The green circles illustrate a sample trajectory.  The uncertainties of individual occupation probability measurements, calculated based on Poissonian counting statistics, are smaller than the markers used. }
\label{fig:compression}
\end{figure}

\paragraph{Results.---} 

The training process is illustrated in Fig.\ \ref{fig:compression}, which shows how the device adapts to the training states over time. The training process was repeated twenty times, each time with a different random initialization of the unitary. We observe that the device was able to achieve a high-quality performance, reducing the occupation probability of the junk output mode to 0.03±0.03. The training was performed with a random but fixed set of two training states and no prior calibration of any wave plates, other than the two half-wave plates fixed at $45^{\circ}$. This demonstrates our device performing in general conditions, with no prior information needed about the unitary transformation or quantum states.

\begin{figure}[tb]
	\centering
	\includegraphics[width=7.5cm]{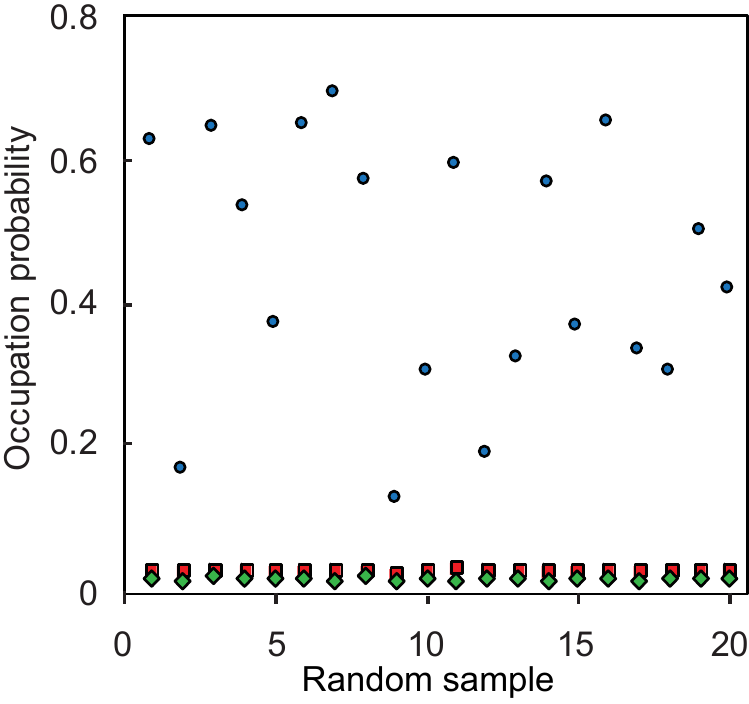}
	\caption{The effect of the number of training states. The optimization routine was run using one, two, and three training states taken from a compressible subspace. After each training process, twenty random states from the same subspace were prepared and sent through the device to test the compression performance of the unitary. The occupation probability of the junk mode after using one training state (blue circles) was 0.4±0.3, with two training states (red squares) it was 0.03±0.02, and with three training states (green diamonds) it was 0.02±0.02. The uncertainties of individual data points, calculated based on Poissonian counting statistics, are smaller than the markers used.}
\label{fig:training}
\end{figure}

The quality of the unitary found by the machine learning process depends on the number of different training states used, as shown in Fig.\ \ref{fig:training}. We explored this relationship by training the unitary with different sized sets of qutrits from a compressible subspace, terminating the training at 200 cost function evaluations, and then testing the device with new, random qutrits from the same subspace. 
Using a single training state, a low occupation probability of the junk mode was obtained for the training state, but randomized test states were unable to pass through with a similarly low probability of exiting the junk mode. This is unsurprising because a single state does not span the qubit subspace. Using two training states, we saw a consistently high level of performance for all of the test states. Repeating this process with three training states, we found that the unitary achieved a slightly lower occupation probability of the junk mode across the test states. However, the total training time increased by approximately 40\% compared to the run time for two training states (about 90 minutes), with only minor benefits to the performance. This is why we used two training states in the other parts of the experiment. Note, however, that the run times in our experiment were limited by wave plate rotations, rather than the \blu calculations of the \blk algorithm, and could be made much faster by using fast switching with electro-optic devices. \blu The choice of feedback routine also affects the training time and compression performance. We chose a gradient descent algorithm as a simple way of achieving a reliable device performance, but alternatives, for instance genetic algorithms, could also be considered. \blk

\begin{figure}[tb]
	\centering
	\includegraphics[width=7.5cm]{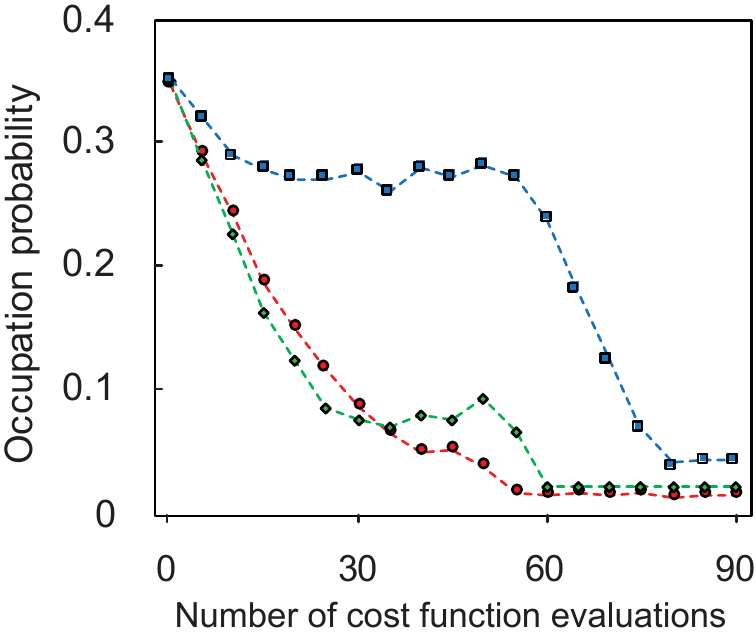}
	\caption{Test of a channel drift during training. A set of two training states was used to optimize the device under normal conditions, providing a control dataset (red circles) of the occupation probability of the junk mode, averaged over the two training states, versus the number of cost function evaluations. For the green and blue datasets, the same unitary initialization was used, but the necessary encoding was perturbed. This was done by keeping the initial qubits in the state preparation fixed but rotating the wave plate of the scrambling stage by 4\textdegree \ per five cost function evaluations. The datasets of blue squares and green diamonds correspond to different rotation directions. The connecting lines in the plot serve merely as a guide to the eye. The uncertainties of individual occupation probability measurements, calculated based on Poissonian counting statistics, are smaller than the markers used.}
\label{fig:drift}
\end{figure}

Finally, we investigated the robustness of the device by introducing a channel drift during the optimization routine, as shown in Fig.\ \ref{fig:drift}. The aim was to emulate a drift in environmental conditions that affect the required encoding. We performed this by systematically rotating the polarization scrambling wave plate in the state preparation by 4\textdegree \ per five cost function evaluations. Despite an increase in the number of required cost function evaluations, the autoencoder was able to achieve an average occupation probability below 0.05. This shows that it is capable of adapting to changes in environmental conditions. \blu The learned transformations are provided in the Supplemental Material \cite[Sec.~D]{SI}.\blk

\paragraph{Discussion.---}

In this work, we have proposed and experimentally demonstrated a new, simple scheme for a quantum autoencoder. We have shown that our device is able to compress qutrits to qubits by exploiting the underlying structure of the dataset. 
The autoencoder is an example of a hybrid quantum-classical machine that optimizes quantum information processing via classical machine learning based on training data. This unsupervised learning provides the valuable capability of adjusting to different datasets, which is absent in alternative methods of compressing quantum data. Indeed, the choice of qubit subspace in our experiment was arbitrary, and was even subjected to a drift during the optimization. 

Comparing our scheme with the proposal of Ref.\ \cite{Romero2017}, we use a different cost function. Our cost function is defined in terms of the occupation probability of the junk mode, which requires a simple measurement of the mode. By contrast, the cost function of Ref.\ \cite{Romero2017} is defined in terms of the overlap between a fixed reference state and part of the output of the encoding unitary, and is estimated via a \textsc{SWAP} test. Although these definitions and measurements appear quite different, they are based on the same principle that no information should be contained in the discarded parts of the state.

Of the encoding-decoding sequence illustrated in Fig. \ref{fig:simple}(a), we have implemented the encoding step. Since this encoding is based on a unitary transformation, the decoding step would simply be the inverse of that transformation \cite{SI}. Furthermore, although our experiment was designed for qutrits, the same design can be extended to compress higher-dimensional qudits. In that case, the number of input modes and the size of the unitary would be increased, \blu with a polynomial scaling in the number of parameters to be optimized \cite{SI}. \blk Depending on the desired output dimensionality, one or more output modes could be monitored. \blu The setup we use can be extended beyond qutrits because of the good stability of beam displacer interferometers and polarization-based two-mode unitary transformations with wave plates \cite{Laing2011,Martin-Lopez2014}. However, an even more promising approach for large unitaries lies in integrated photonics, where a reconfigurable unitary transformation in six dimensions has been demonstrated \cite{Carolan2015}. Indeed, a \blk recent optical implementation of machine learning for a classical application, vowel recognition, has already shown the feasibility of training unitaries in a photonic waveguide architecture \cite{Shen2017}.

To assess the performance of our device, we have focused on lossless compression, for which we prepared families of qutrits that are in principle perfectly compressible into qubits. The opportunity for lossless compression can arise in certain situations, for example when a physical symmetry restricts quantum states to a subspace of the full Hilbert space \cite{Romero2017}, or when a signal originates from a lower-dimensional encoding, but is spread over a larger Hilbert space through an imperfect transmission channel. However, compression can also be of interest in cases where it is impossible to compress the input data completely faithfully. For instance, it might be worth sacrificing the ability to recover the exact input states for the sake of reducing quantum memory requirements. Exploring such irreversible compression with the autoencoder presents an interesting future research direction.

\paragraph{Acknowledgments.---}
We thank Anthony Laing for ideas on designing unitaries for qutrits, Nana Liu and Lucas Lamata for useful discussions, and Raj Patel for contributions of data acquisition code. This work was supported by the Australian Research Council Grant No.\ DP160101911, and by a Griffith University New Researcher Grant. We acknowledge the traditional owners of the land on which this work was undertaken at Griffith University, the Yuggera people.
%\onecolumngrid

%\twocolumngrid
%\bibliographystyle{abbrv}
%\bibliographystyle{unsrt}
\bibliography{AutoencoderRefs}
%\bibliography{withfootnotes}

\onecolumngrid

\newpage 

%%%%%%%%%%%%%%%%%%%%%%%%%%%%%%%%%%%%%%%%%%%%% SM starts here %%%%%%

\setcounter{equation}{0}
\renewcommand{\theequation}{S.\arabic{equation}}

\setcounter{figure}{0}
\renewcommand{\thefigure}{S\arabic{figure}}

\renewcommand{\thepage}{Supplemental Material -- \arabic{page}/2}
\setcounter{page}{1}

\begin{center}
  \large {\bf Supplemental Material for \red ``Experimental Realization of \blk a Quantum Autoencoder:\\  The Compression of Qutrits via Machine Learning''}
\end{center}
\vskip 1em
\begin{center}
   \lineskip .75em%
  \begin{tabular}[t]{c}
    Alex Pepper, Nora Tischler, and Geoff J. Pryde
  \end{tabular}%\par
\end{center}
\section{A: Generalization to higher dimensions}

\begin{figure}[h]
	\centering
	\includegraphics[width=16.5cm]{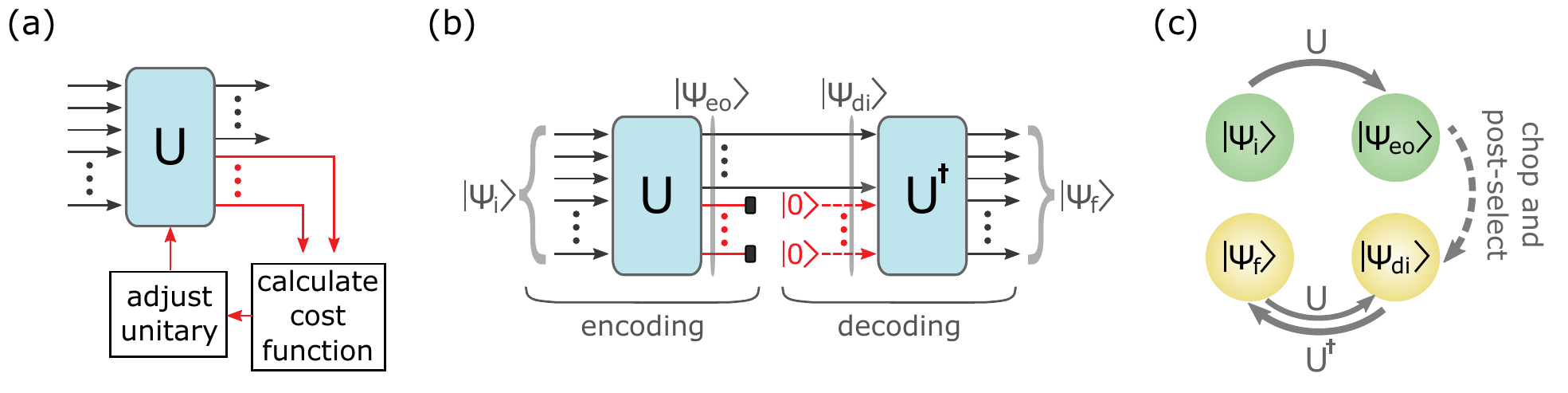}
	\caption{(a) Training of an autoencoder that reduces qudits to qunits. The cost function is calculated based on the occupation probabilities of the $(d-n)$ junk modes. (b) The encoding-decoding sequence for the autoencoder. Ideally, the decoded state $|\Psi_f\rangle$ is equal to the input state $|\Psi_i\rangle$ for any member of the dataset. (c) Relationship between different quantum states in the encoding-decoding sequence. $|\Psi_{\mathrm{i}}\rangle$ is an input state of the autoencoder; $|\Psi_{\mathrm{eo}}\rangle$ is the output of the encoding unitary $U$; $|\Psi_{\mathrm{di}}\rangle$ is the result of discarding the junk modes and replacing them with vacuum, post-selected on the photon being in the first $n$ modes after the encoding unitary; $|\Psi_{\mathrm{f}}\rangle$ is the output of the decoding transformation $U^{\dagger}$.}
\label{fig:SF}
\end{figure}

\noindent The scheme can be used more generally to reduce a dataset of $d$-dimensional qudits to $n$-dimensional qunits, as illustrated in Fig.\ \ref{fig:SF}(a). At the output of the $(d\times d)$-dimensional encoding unitary $U$, the first $n$ modes are retained as the modes of the encoded states, while the last $(d-n)$ modes constitute the junk modes. \blu The unitary $U$ contains $\left[ d(d-1)/2 - n(n-1)/2 \right]$ two-mode unitaries, each of which entails two free parameters to be optimized. This amounts to fewer free parameters than for a general $d$-dimensional unitary \cite{Reck1994,Laing2011,Martin-Lopez2014}, since the single-mode phase shifters and $ \left[ n(n-1)/2\right]$ two-mode unitaries that act within the subspace of the first $n$ modes are unnecessary. \blk Denoting the occupation probability of the $k$-th junk mode for the $m$-th training state by $P^m_{\mathrm{j}k}$, \blu and letting $m_\mathrm{max}$ be the number of training states, \blk the cost function for training the unitary transformation can be defined as \blu $C=\sum_{m=1}^{m_\mathrm{max}} \sum_{k=1}^{d-n} P^m_{\mathrm{j}k}/m_\mathrm{max}$ \blk (other definitions are also possible). 

After encoding quantum states from the dataset with a trained unitary $U$, they can be decoded by acting with $U^{\dagger}$ on the encoded states, after augmenting them with $(d-n)$ vacuum inputs modes (see Fig.\ \ref{fig:SF}(b)). 

\section{B: Quantifying the device performance}
\noindent In our device, two types of error need be minimized for any given input state: (i) the probability that the photon gets discarded due to it occupying the junk modes, in which case the autoencoder fails to produce an output photon, and (ii) the difference between the original input, $|\Psi_{\mathrm{i}}\rangle$, and the output after an encoding-decoding sequence, $|\Psi_{\mathrm{f}}\rangle$. We use the probability that the photon occupies one of the junk modes, $P_{\mathrm{j}}= \sum_{k=1}^{d-n} P_{\mathrm{j}k}$, as a measure of both types of error. 

Clearly, the probability that the photon gets discarded due to it occupying the junk modes is simply $P_{\mathrm{j}}$. In other words, the success probability of the autoencoder producing an output is $1-P_{\mathrm{j}}$.

Given that an output is produced, the fidelity of the input and output of the encoding-decoding sequence is also determined by $P_{\mathrm{j}}$. This can be seen as follows: Let $\{\alpha_{n_1} |~ n_1\in \{1,2,...,n \}\}$ be the complex amplitudes of the first $n$ modes of $|\Psi_{\mathrm{eo}}\rangle$ and $\{\beta_{n_2} | ~n_2\in \{1,2,...,d-n \}\}$ the amplitudes of the junk modes, such that $|\Psi_{\mathrm{eo}}\rangle=(\alpha_1, ..., \alpha_{n},\beta_1, ..., \beta_{d-n})^T$. Then $|\Psi_{\mathrm{di}}\rangle=(1-P_{\mathrm{j}})^{-1/2}(\alpha_1, ..., \alpha_{n},0, ..., 0)^T$, and
\begin{eqnarray}
|\langle \Psi_{\mathrm{i}}|\Psi_{\mathrm{f}} \rangle|^2&=&|\langle \Psi_{\mathrm{eo}}|\Psi_{\mathrm{di}} \rangle|^2\nonumber \\
&=&|(1-P_{\mathrm{j}})^{-1/2}(\alpha_{1}^*\alpha_{1}+...+\alpha_{n}^*\alpha_{n})|^2\nonumber \\
&=&1-P_{\mathrm{j}}.
\end{eqnarray}

\blu
\section{C: Training algorithm}
\noindent The gradient descent algorithm we use for training the autoencoder is an iterative procedure that aims to find a minimum of the cost function $C(\mathbf{x})$, which depends on the four motorized wave plate angles in the autoencoder, $\mathbf{x}=\left( x_1,x_2,x_3,x_4\right) $. At the start of the training procedure, the angles are initialized randomly. Each iteration consists of the two stages described below: 1.\ a probing stage and 2.\ a movement stage. 

\begin{enumerate}
\item The purpose of the probing stage is to estimate an approximate gradient, $\nabla C|_{\mathbf{x_\mathrm{cur}}}$, at the current wave plate settings $\mathbf{x}=\mathbf{x_\mathrm{cur}}$ through discretization. In order to estimate the partial derivative with respect to the $n$-th variable, the $n$-th wave plate is individually rotated by an adjustment value, $s_a$, resulting in a configuration $\mathbf{x}_{\mathrm{p}n}$. There, the cost function is measured, which requires a sequential preparation of the different training states, before returning the wave plate to its previous orientation. The partial derivative with respect to the $n$-th wave plate angle is then estimated as the slope of the secant line $ \left. \frac{\partial C}{\partial x_n}\right|_{\mathbf{x_\mathrm{cur}}}=\left[ C(\mathbf{x}_{\mathrm{p}n})-C(\mathbf{x_{\mathrm{cur}}})\right]/ s_a $. 
\item	In the movement stage, a simultaneous rotation of all four wave plates with a step size proportional to $s_a$ is made in the opposite direction of the gradient: $\mathbf{x_\mathrm{cur}} \rightarrow \mathbf{x_\mathrm{cur}}-s_a \nabla C|_{\mathbf{x_\mathrm{cur}}}$. 
\end{enumerate}

\noindent Two different adjustment values, a coarse value and a fine value, are used at different times of the training procedure. The coarse adjustment value of $s_a= 12^{\circ}$ is used to quickly approach a minimum at the beginning of the training procedure. Once the cost function crosses a threshold value of 0.1, a smaller value of $s_a= 5^{\circ}$ is used for increased precision. Note that although we use $s_a$ from stage 1 again in stage 2, in general, a different proportionality coefficient for the step size (learning rate) could also have been used in stage 2.

In addition to the above, a deliberate disturbance is incorporated if a specific condition occurs: If the device is unable to achieve a cost function below 0.1 within 50 steps, each wave plate is rotated by 25°. The purpose of this is to bump the device out of poor optimizations and allow the reconfiguration to more desirable values. The conditional disturbance was added based on earlier observations of the training procedure, where for some of the training runs the cost function plateaued at values of 0.15±0.05. This extra feature serves to ensure a reliable device performance.
\blk

\blu
\section{D: Example transformations}
\noindent Here, we provide the three unitary transformations which the autoencoder learned in the training runs depicted in Fig.\ \ref{fig:drift}. The unitaries are calculated based on a model of the setup and the experimental configuration of the wave plates at the end of the training runs. We denote the transformation learned in the control run as $U_1$, and the two transformations learned under perturbations as $U_2$ and $U_3$. The transformations are
\[
U_1
=
\begin{bmatrix}
    -0.373 - 0.037 i & -0.927 + 0.015 i & 0  \\
   0.008 + 0.213 i & -0.013 - 0.085 i & -0.003 - 0.973 i  \\
    -0.017 + 0.902 i & -0.035 - 0.363 i & -0.012 + 0.230 i 
\end{bmatrix},
\]

\[
U_2
=
\begin{bmatrix}
   0.054 - 0.610 i & -0.790 - 0.042 i & 0  \\
   -0.090 + 0.094 i & -0.075 - 0.067 i & 0.014 - 0.986 i  \\
    0.052 + 0.778 i & -0.601 + 0.062 i & 0.120 + 0.112 i 
\end{bmatrix},
\]

\[
U_3
=
\begin{bmatrix}
   0.495 + 0.298 i & -0.795 + 0.185 i & 0  \\
  -0.172 + 0.228 i & 0.025 + 0.200 i & 0.048 - 0.936 i  \\
   -0.135 + 0.753 i & 0.303 + 0.449 i & 0.142 + 0.319 i 
\end{bmatrix}.
\]

\noindent Note that the $(1,3)$ element of each unitary is 0 because the autoencoder does not comprise, or require, a fully general unitary transformation between the three modes. Our autoencoder can omit one of the $2\times 2$ transformations of a general decomposition \cite{Reck1994,Laing2011,Martin-Lopez2014}, namely the one in the encoded output subspace, without affecting the cost function or the ability to learn how to compress any compressible set of qutrits.
\blk

\end{document}